\documentclass[12pt,preprint]{aastex}

\begin{document} 

\title{THE NON-HOMOLOGOUS NATURE OF SOLAR DIAMETER VARIATIONS} 

\author{Sabatino Sofia\altaffilmark{1}, Sarbani Basu\altaffilmark{1}, Pierre 
Demarque\altaffilmark{1}, Linghuai Li\altaffilmark{1}, and Gerard 
Thuillier\altaffilmark{2}} 
\altaffiltext{1}{Department of Astronomy, Yale University, P.O. Box 208101, New 
Haven, CT 06520-8101} 
\email{sofia@astro.yale.edu} 
\email{basu@astro.yale.edu} 
\email{demarque@astro.yale.edu} 
\email{li@astro.yale.edu} 
\email{gerard.thuillier@aerov.jussieu.fr} 
\altaffiltext{2}{Service d'A\'{e}ronomie du CNRS, Bp 3, 91371 Verri\`{e}res-le 
Buisson, France} 

\begin{abstract} 
We show in this paper that the changes of the solar diameter in response 
to variations of large scale magnetic fields and turbulence are not homologous. 
For the best current model, the variation at the photospheric level is over 
1000 times larger than the variation at a depth of 5 Mm, which is about the level 
at which f-mode solar oscillations determine diameter variations. 
This model is supported by observations that indicate larger diameter 
changes for high degree f-modes than for low degree f-modes, since energy of the 
former are concentrated at shallower layers than the latter.   
\end{abstract} 
\keywords{Solar diameter: variations} 

\section{Introduction} 

The question of whether the solar diameter changes on timescales of years to 
centuries is very controversial.  A recent paper (Thuillier et al 2005a) 
presents a detailed summary of the issue.  In essence, different measurement 
and analysis techniques, and sometimes even identical instruments and 
similar analysis methods, yield incompatible results.  It is not presumptuous 
to infer that the cause of this controversy is that for the majority of 
the techniques, the results are at the borderline of the 
sensitivity of the technique. 

The major exception to the above statement is the technique of 
helioseismology, particularly of the f-modes of oscillation. 
Schou et al. (1997) and Antia (1998) have demonstrated that the 
frequencies of f-modes can be used to estimate the solar radius. 
Since these frequencies have been measured with a precision of one 
part in $10^5$, one may expect to determine the solar radius to 
similar precision. 
Changes in the f-mode frequencies have been used 
to determine changes in the solar radius (see e.g., Dziembowski et al. 1998, 
Antia et al. 2000, etc.). The radius changes are estimated 
assuming that the fractional change in radius is uniform in the 
range of sensitivity of the method. 
The radius change determined by 
f-modes is the change at the radius where the f-modes are concentrated. One 
way of quantifying the depth at which the f-modes are sensitive 
is to look at the depth at which energy of the f-modes is concentrated. 
This is shown in Fig.~1, where we plot the density for f-modes of 
several degrees. 
The range of degrees on the plot reflects the range of available f-mode frequencies. 
We see that for the lowest degree mode in the figure ($\ell=140$), the 
peak in the energy is at about 6.3 Mm (where temperature 
$T$ is about 41000K), and for the highest degree ($\ell=300$) it 
is at 3.3 Mm ($T=24000$K). 
There are of course, other criteria by which 
one can determine radius  (see e.g. Cox 1980, Unno et al. 1989). However, here we 
show the energy density because that is the criterion used by various authors when 
discussing their radius-change results. In any event, the different methods do not significantly
alter the conclusions of this paper.
It should be noted that although the f-modes 
have a precision of one part in $10^5$ or so, the 
changes in radius cannot be determined with this precision since that 
depends on how large the changes in f-modes are, and these changes 
happen to be very small. As a result,  the radius-change 
measurements are not always very precise as can be seen from the 
results shown below. 

Consequently, even in the case of radius determination using f-mode oscillations, there does not 
seem to be consensus yet as to the exact amount by which the 
radius changes. The results obtained so far (Dziembowski et al. 
1998, 2000, 2001; Antia et al. 2000, 2001; Antia \& Basu 2004) 
are not in agreement with each other. Dziembowski et al. 
(1998), using MDI data, found that the solar radius reached a 
minimum around the minimum activity period in 1996 and 
was larger, by about 5 km, 6 months before and after the minimum. 
However, later results, using longer time intervals, did not find 
any systematic changes (Dziembowski et al. 2000). On the 
other hand, Antia et al. (2000) using Global Oscillation Network 
Group (GONG) data found that the solar radius decreased by about 
5 km between 1995 and 1998, and this variation appeared to be 
correlated (but in antiphase) with the level of solar activity. 
Subsequently Antia et al. (2001), using both GONG and MDI data, put 
an upper limit of 1 km/year for the change in solar radius. Meanwhile, 
on a related paper, Dziembowski et al. (2001) claimed a solar 
radius decrease as a rate of 1.5 km yr$^{-1}$ during 1996-2000. 
Antia et al. (2001) made some sense of all these discrepant results 
by showing that the variation in f-mode frequencies could be 
divided into at least two components: one oscillatory, with a 
period of 1 yr, and a second, non-oscillatory, and probably 
correlated with solar activity. They argued that the oscillatory 
component is most likely an artifact introduced 
by the orbital period of the Earth. They also 
showed that most of the discrepancy between different results 
could be explained by the use of data sets that cover different 
time periods and by the failure to remove the oscillatory component. 
Upon performing those corrections, all the different investigations 
appear to indicate that the solar radius decreases with increasing 
solar activity. 

In a more recent investigation  Antia \& Basu (2004) examined the 
changes in f-mode frequencies using eight years of MDI data. They 
obtain an upper limit of about 1 km/year for radius changes during 
the entire solar cycle. It is to be noted, however, that even this 
result is not very clear cut, since different degree ranges of 
f-modes implied different radius changes. When the available 
higher degree modes were used ($ 140 < \ell \le 300$), they got an 
average change of $-0.91\pm0.03$ km yr$^{-1}$ between 1996 and 
2004. F-modes in the range $140\le\ell\le 250$ show a change of 
$-0.41\pm 0.04$ km yr$^{-1}$ for the same period, but for 
$\ell < 140$, no observable change was obtained, 
($\Delta R =0.13\pm 0.20$ km per year). Antia \& Basu (2004) suggested 
that the difference in the results yielded by the different 
degree ranges indicated that the evidence for radius change 
was not conclusive.  In this paper we will present an alternative 
interpretation of these observations, i.e., that the Sun does not  expand or 
contract homologously with the change in solar activity.

\section{Model calculations} 

We construct models to calculate the change in solar radius with change in solar 
activity. 
The numerical code that we use to compute the structure and evolution of 
the solar model is an outgrowth of YREC (Winnick et al 2002) into which the effects of magnetic 
fields and turbulence have been included. The starting values of the basic 
solar papameters are:  $R_\sun  = 6.9598 \times 10^{10}$ cm 
and $L_\sun = 3.8515\times 10^{33}$ erg/s.
These particular choices have negligible effects on the results.
The version of the code used in 
these calculations is one-dimensional. The inclusion of magnetic fields 
considers their contribution to pressure and internal energy, and 
their modification of energy transfer, primarily convection.  The dynamical 
effects modify turbulent pressure and energy transport.  The detailed 
formulation of the modifications to YREC is based on the approach first 
presented by Lydon \& Sofia (1995), and subsequently expanded by Li \& Sofia 
(2001), and Li et al. (2002, 2003). 

Because the location, magnitude and temporal behavior of the internal field 
are not known, we made two general assumptions: (1) the magnitude of the 
magnetic field would be that required to cause a luminosity change of 0.1 
percent over the cycle, and (2) the temporal behavior assumed is 
sinusoidal, and it mimics the shape of 
the activity cycle determined, for example, by the averaged sunspot number. 
We computed four cases (listed in Table 1), three with only magnetic fields at 
different depths, and one with both magnetic fields and turbulence.  For 
the cases with only magnetic fields, the field configuration was Gaussian.   
Guided by the observation of p-mode oscillations, we were led to the 
inclusion of turbulence (Li et al. 2003).  In this case, the 
properties of the turbulence were 
derived from numerical simulations of the outer region of the solar convective 
envelope (Robinson et al. 2003), and the magnetic field distribution was 
dictated by a feedback process between turbulence and magnetic field. 

Although the specific details of the calculations reported here are contained 
in Li et al. (2003) 
we present in this paper the results of the calculations that are relevant to 
the radius problem, and not contained 
in that paper. The lower panel of Fig.~2 shows the difference in magnetic 
energy per unit mass ($\chi_m\equiv B^2/8\pi\rho$) 
between the years 2000 and 1996 for all cases where only magnetic fields 
are taken into account.  The upper panel of the figure shows the ratio between 
the radius change as a function of radius to the radius change at 5 Mm. 

Fig.~3 is similar to Fig.~2, but for the case in which turbulence (modulated 
by the magnetic field) is included. The lower panel shows the difference 
in turbulent ($\chi_t\equiv \frac{1}{2}(v'')^2$, where $v''$ is the magnitude of the turbulent velocity) plus 
magnetic ($\chi_m$) energy per unit mass between the years 2000 and 1996 ($\chi=\chi_m+\chi_t$), 
and the upper panel 
the ratio between the radius change as a function of radius to the radius 
change at 5 Mm. 

From Fig.~2 we notice that in all cases the radius increases with increasing 
solar activity.  This is to be expected, since all the contribution of 
the magnetic field to pressure and internal energy is positive, and 
consequently, it can only lead to an increase of the radius.  We also 
notice that the increase of the radius is monotonic towards the surface. 
This is because the increase at a given radius is made up of the sum of 
the increase at all levels below it.  Finally, we notice that the 
expansion, which only increases in the magnetic region, is accelerated 
towards the shallower layers. This can be understood since, for a given 
value of the magnetic field, the ratio of magnetic to total pressure increases 
with increasing radius, and so does the expansion. 

Fig.~3 represents the case that, according to Li et al. (2003), meets all the 
observational requirements imposed by helioseismology.  In particular, 
it produces the 
correct cycle-related variations of the p-mode oscillations, it does not 
alter the depth of the convection zone, and it produces diameter changes 
in opposite phase of the activity cycle.  In this case, the magnetic field 
slows down turbulent flows so that the increase of magnetic pressure when 
the magnetic field grows is overcompensated by the corresponding decrease in 
turbulent pressure. 

In all cases, the radius variation at the solar surface [which is measured 
by any limb-observing instrument, such as the Solar Disk Sextant (Sofia et al. 
1994), 
and PICARD Thuillier et al. (2005b)], 
can be hundreds of times larger than the radius variation inferred by 
f-mode oscillations, which represents changes at several Mm. 

To determine the depth of the level of the ``diameter'' provided by the 
f-mode oscillations, we refer to Fig. 1, which represents the kinetic 
energy of the modes of different $\ell$-values (abscissa in arbitrary units). 
We can see that for higher $\ell$s, the peak energy occurs at shallower layers 
than for lower-$\ell$ modes.  It would appear that 5 Mm is a good number to 
represent the depth of the layer given by f-mode oscillations of all degrees 
observed.  Thus, the upper panel of Figs. 2 and 3 give the magnification 
factor between radius changes determined from f-mode oscillations, and the 
radius changes that can be expected at the photospheric level, and thus 
to be observed by all limb-observing instruments. 

Because $\Delta R$ increases in the shallower layers, the high-$\ell$ modes, which 
peak in shallower layers, should show a larger radius change than the 
low-$\ell$ modes, which peak at deeper layers.  We believe this is precisely what the 
results obtained by Antia \& Basu (2004) imply.

\section{Summary and Conclusions} 

We have shown that the model of variability of the solar interior that   
obeys all observational constraints (global parameters and p-mode and 
f-mode oscillations) produces variations of the solar radius that 
increase by a factor of approximately 1000 from a depth of 5 Mm to the 
solar surface.  This model includes the effects of a variable dynamo 
magnetic field, and of a field-modulated turbulence, and it explains features 
of the f-mode oscillations in different degree ranges that were 
previously not understood. 

On the basis of the above argument, we conclude that results from f-mode 
oscillations that the solar radius only changes by about 1 km/year does 
not preclude the less-sensitive efforts to measure variations of the 
solar radius at the photosphere by limb observations since the latter 
are likely to be much larger than the former. Limb observations are 
made by a number of ground-based instruments, and from above the atmosphere 
by the Solar Disk Sextant (SDS) balloon-borne experiment (Sofia et al. 1994), and 
will be made starting in 2008 by the PICARD microsatellite (Thuillier et al. 
2005b). In thermal equilibrium the space-based instruments have a theoretical 
precision of about 1 milli arc s (about 1 km). PICARD should easily 
reach such a precision.  Balloon based observations, however, cannot 
reach this precision because the short duration of the flights prevents the 
instrument from reaching thermal equilibrium.  The current SDS results have   
reached a precision of the order $\pm 0.05$ arc s (Egidi et al 2005). 

\acknowledgments 
This work was supported in part by 
NSF grants ATM 0206130 and ATM 0348837 to SB. SS and PD were supported in part by NASA grant NAG5-13299.

\newpage 

\begin{deluxetable}{ccccc} 
\tablecaption{Solar variability models. \label{tbl:model0}} 
\tablewidth{0pt} 
\tablehead{ 
 Model & Depth & $\Delta B$ & $P_m/P$ & $P_t/P$ \\ 
 &  (Mm) & (kG) & (\%) & (\%) } 
\startdata 
 1 & 4.45   & 4.2  & 0.66 & 0 \\ 
 2 & 2.43   & 1.7  & 1.1  & 0 \\ 
 3 & 0.38   & 0.15 & 0.67 & 0 \\ 
 4 & 2.33   & 0.38 & 1.2  & 16\\ 
\enddata 
\end{deluxetable} 

\begin{figure} 
\plotone{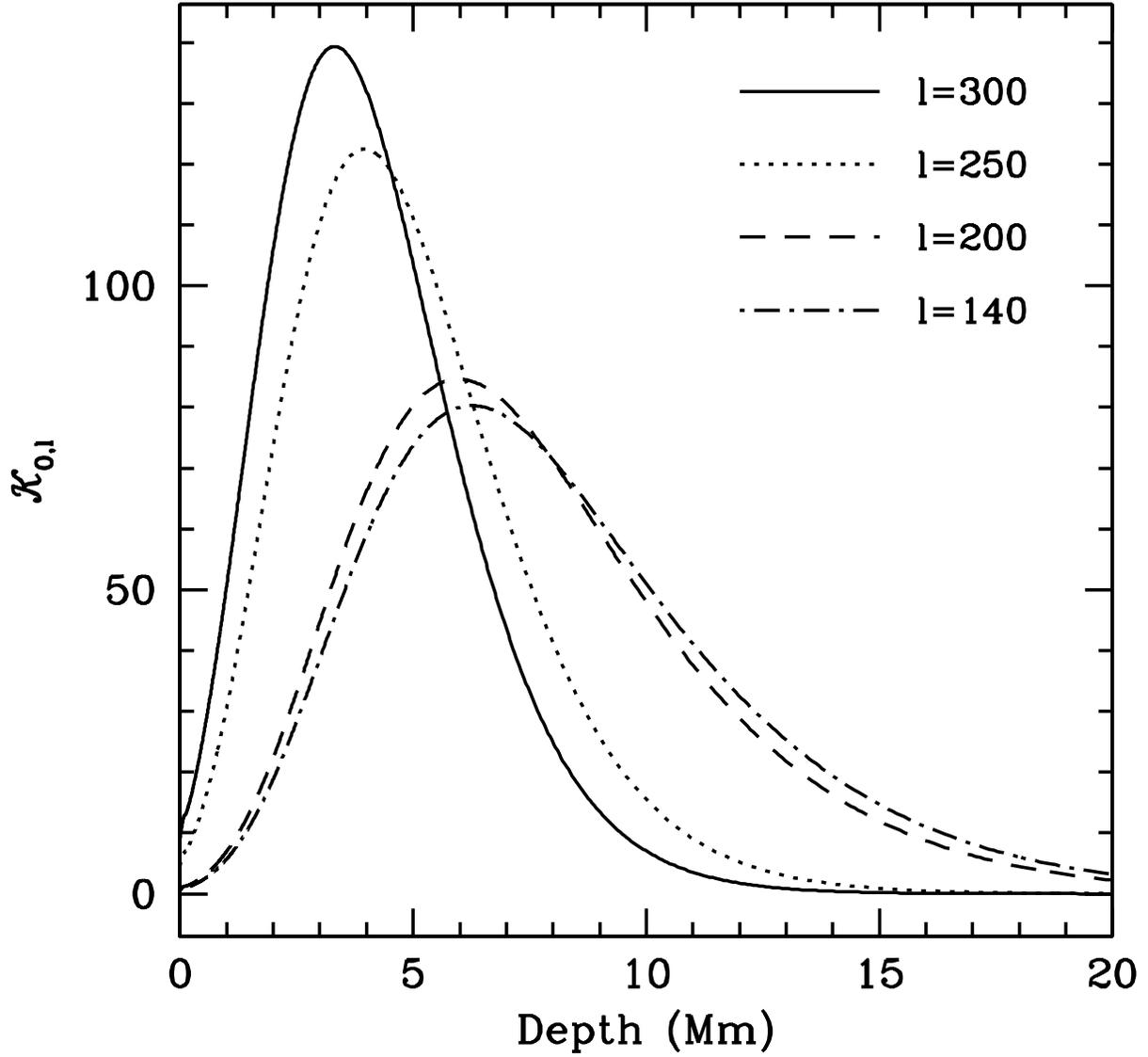} 
\caption{ 
Kinetic energy density of different f-modes. The curves are 
normalized by the total kinetic energy density in each mode. The 
eigenfunction used to calculate the energy density were those of the 
standard solar model of Basu, Pinsonneault and Bahcall (2000). 
\label{fig:f1}} 
\end{figure} 

\begin{figure} 
\plotone{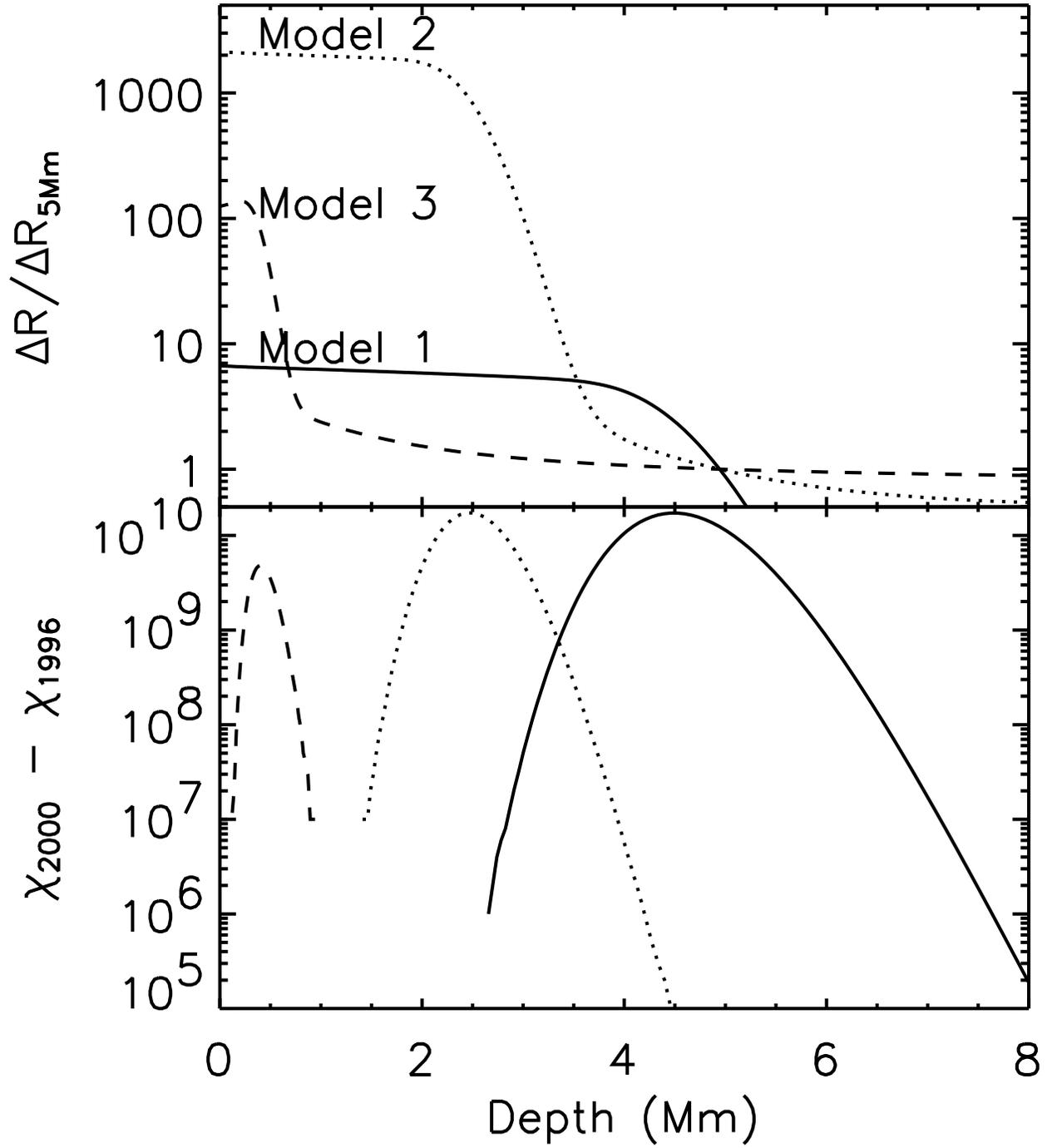} 
\caption{ 
Lower panel: Magnetic energy density for models 1 (solid line), 2 (dotted 
line) and 3 (dashed line).  The model properties are listed in Table 1. 
Upper panel: The ratio between radius change as a function of depth below 
the photosphere and radius change at 5 Mm for models 1-3. 
\label{fig:f2}} 
\end{figure} 

\begin{figure} 
\plotone{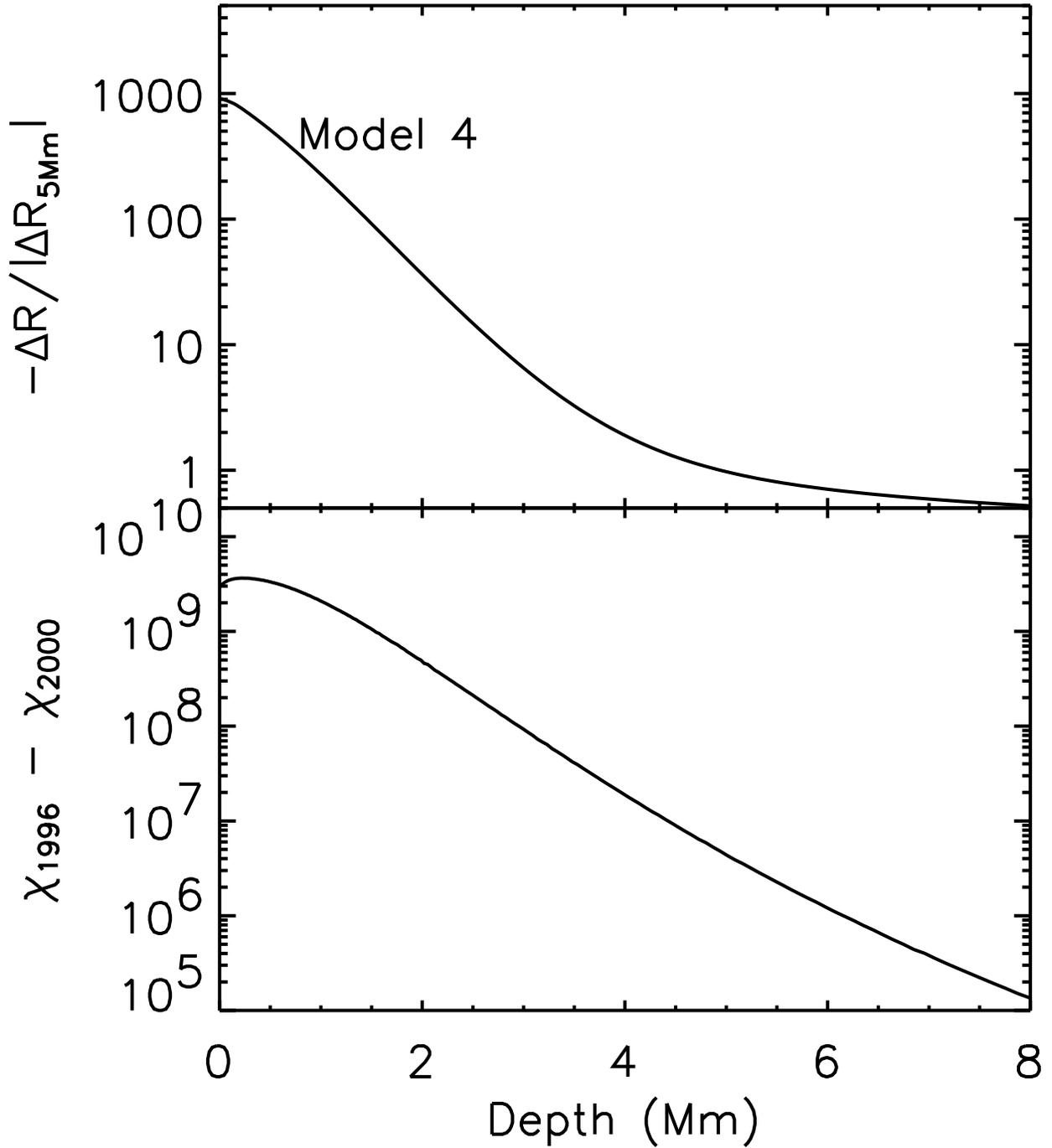} 
\caption{ 
Lower panel: Magnetic plus turbulent energy density for model 4, where 
turbulence is modulated by the magnetic field.  Notice that the increase 
in magnetic field density between 1996 and 2000 is more than offset by 
the decrease in turbulent energy density, so that the total energy density 
decreases with increasing level of activity. 
Upper panel: The ratio between radius change as a function of depth 
below the photosphere and radius change at 5 Mm for model 4.  Notice 
that the radius decreases with increasing solar activity.   
\label{fig:f3}} 
\end{figure}

\end{document}